\documentclass[prl,twocolumn,notitlepage,showpacs,amsmath,amstex,amssymb,citeautoscript,longbibliography,superscriptaddress,groupedaddress]{revtex4-1}
\pdfoutput=1

\usepackage{natbib}
\usepackage[english]{babel}
\usepackage{letltxmacro}
\usepackage{latexsym}
\LetLtxMacro{\ORIGselectlanguage}{\selectlanguage}
\makeatletter
\DeclareRobustCommand{\selectlanguage}[1]{%
  \@ifundefined{alias@\string#1}
    {\ORIGselectlanguage{#1}}
    {\begingroup\edef\x{\endgroup
       \noexpand\ORIGselectlanguage{\@nameuse{alias@#1}}}\x}%
}
\newcommand{\definelanguagealias}[2]{%
  \@namedef{alias@#1}{#2}%
}
\makeatother
\definelanguagealias{en}{english}
\definelanguagealias{English}{english}
\usepackage{graphicx}
\usepackage{amsmath}
\usepackage{mathtools}
\usepackage{amsfonts}
\usepackage{amssymb,bbding}
\usepackage{bm}
\usepackage{color}
\usepackage[percent]{overpic}
\usepackage{soul} 
\usepackage{amssymb}
\usepackage{wasysym}
\usepackage{dsfont}
\usepackage{float}
\usepackage{braket}
\usepackage{subfiles}
\usepackage{epstopdf}

\usepackage[normalem]{ulem}

\usepackage{tikz}
\usetikzlibrary{positioning}

\usepackage{blkarray}
\usepackage{multirow}

\usepackage{hyperref}
\hypersetup{
    bookmarks=false,         
    unicode=false,          
    pdftoolbar=false,        
    pdfmenubar=true,        
    pdffitwindow=false,     
    pdfstartview={FitH},    
    pdftitle={Physical insights into approximating quantum many-body states with machine learning ansatz},    
    pdfauthor={A. Borin, D. Abanin},     
    pdfsubject={},   
    pdfcreator={},   
    pdfproducer={}, 
    pdfnewwindow=true,      
    colorlinks=true,       
    linkcolor=black,          
    citecolor=blue,        
    filecolor=magenta,      
    urlcolor=blue           
}

\newcommand{\be}{\begin{equation}}
\newcommand{\ee}{\end{equation}}
\newcommand{\bea}{\begin{eqnarray}}
\newcommand{\eea}{\end{eqnarray}}

\begin{document}

\title{Approximating power of machine-learning ansatz for quantum many-body states}

\author{Artem Borin and Dmitry A.\ Abanin}
\affiliation{
 D\'epartement de Physique Th\'eorique, Universit\'e de Gen\`eve - CH-1211 Gen\`eve 4, Switzerland}

\date{\today}

\begin{abstract}
An artificial neural network (ANN) with the restricted Boltzmann machine (RBM) architecture was recently proposed as a versatile variational quantum many-body wave function. 
In this work we provide physical insights into the performance of this ansatz. We uncover the connection between the structure of RBM and perturbation series, which explains the excellent precision achieved by RBM  ansazt in certain simple models, demonstrated in Ref.~[\onlinecite{carleo}]. Based on this relation, we improve the numerical algorithm to achieve better performance of RBM in cases where local minima complicate the convergence to the global one. We introduce other classes of variational wave-functions, which are also capable of reproducing the perturbative structure, and show that their performance is comparable to that of RBM. Furthermore, we study the performance of a few-layer RBM for approximating ground states of random, translationally-invariant models in 1d, as well as random matrix-product states (MPS). We find that the error in approximating such states exhibits a broad distribution, and is largely determined by the entanglement properties of the targeted state.
\end{abstract}

\maketitle
\nocite{carleo}

{\it Introduction.}--- Variational methods play an invaluable role in quantum many-body physics, because they allow one to represent exponentially many amplitudes of a many-body wave function using a small number of variational parameters. The choice of the variational ansatz is often motivated by the underlying physics of the system of interest, notable examples being product states, BCS wave function~\citep{BCS}, and Laughlin states~\citep{Laughlin}. Broad classes of variational wave functions, such as tensor networks~\citep{tensor,verstraete,peps}, which include matrix product states (MPS)~\citep{verstraete}, rely on the low amount of quantum entanglement in ground states of physical systems. 

\begin{figure}[t]
		\includegraphics[width=0.95\columnwidth]{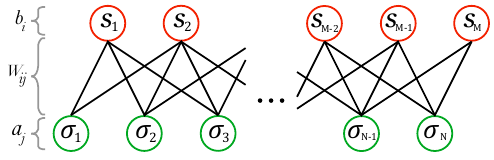} 
	\caption{Illustration of the RBM architecture: $\{ \sigma_j^z \}$ denote physical spins, and $\{s_i\}$ are the hidden spins. The variational parameters $\{a_j\}$, $\{b_i\}$ represent biases applied to physical and hidden spins, and parameters $\{W_{ij}\}$ correspond to couplings between the two flavors of spins.}
\label{fig:RBM_struct}
\end{figure}

Even though tensor network methods proved remarkably successful~\citep{dmrg,dmrg2,tebd,peps,mera}, it is important to investigate other classes of variational wave functions, including the ones that can capture states with higher entanglement. Recent proposals~\citep{carleo,app1,app2,app3,app4,app5,app6,app7,app8,app9,app10} for such ans\"atze considered variational functions inspired by machine learning (ML). More generally, ML is finding an increasing number of diverse applications in physics, including detection of phase transitions~\citep{phas1,phas2,phas3,phas4,phas5,phas6,phas7,phas8,phas9,phas10}, extraction of relevant degrees of freedom~\citep{sampcl,mutual}, and improvement of existing techniques \citep{MC1,MC2,tomog}. In this paper we consider the variational ansatz inspired by the restricted Boltzmann machine (RBM)~\citep{carleo}. The architecture of RBM -- a neural network with a wide range of applications outside quantum physics~\citep{rbm1,rbm2,rbm3} -- is illustrated in Fig.\ \ref{fig:RBM_struct}.

The RBM ansatz for the system of $N$ physical spins $\{\sigma_i^z\}=\pm 1$ is constructed by introducing $M$ hidden spins $\{s_i\}=\pm 1$, which gives $\alpha = M/N$ hidden spins per each physical spin. Then, the amplitudes of the variational wave function in the $\{\sigma^z_i\}$ eigenbasis are obtained by summing over the states of hidden degrees of freedom:
\begin{equation}
\label{RBM_orig}
\Psi_{RBM}(\{\sigma_j\}) = \sum_{\{s_i = \pm 1\}} e^{\sum_j a_j \sigma^z_j +\sum_i b_i s_i + \sum_{ij} W_{ij}s_i\sigma^z_j}, 
\end{equation}
This wave function is optimised over the variational parameters $\{a_j\}$, $\{b_i\}$, $\{W_{ij}\}$ to yield the lowest-energy wave function for a given Hamiltonian.  

The summation over the hidden-spin states in Eq.\ \eqref{RBM_orig} can be performed explicitly, giving (up to a normalization factor):
\begin{align}
\Psi_{RBM}(\{\sigma_j\}) \propto \; &\exp \left({\sum_j a_j \sigma^z_j} \right)\times\nonumber\\
&\prod^M_{i=1}\cosh\left(b_i+\sum_{j}W_{ij}\sigma^z_j\right).
\label{RBM_ans}
\end{align}
This property facilitates Monte Carlo sampling, since any quantum amplitude can be computed by evaluating the function \eqref{RBM_ans}.  

The utility of the RBM ansatz \eqref{RBM_ans} is being actively explored~\citep{Entanglement,topo_st,MPStoRBM,ut1,ut2,ut3}. Among its advantages are applicability in any number of spatial dimensions, and the ability to describe states with high degree of entanglement. It was found to approximate the ground states (GS) of certain Hamiltonians (1D Ising, 1D and 2D Heisenberg)~\citep{carleo} with a remarkably high precision, given relatively few variational parameters. Moreover, it can exactly represent certain topological states (1D cluster state, 2D and 3D toric code)~\citep{topo_st}. 

An important question regarding the representation power of tensor-network states by RBM ansatz is intimately related to the entanglement properties of the latter. 
The entanglement entropy of a general RBM state obeys volume-law scaling~\citep{Entanglement}, but finite-range RBM states (that is, states in which a hidden spin can have non-zero interaction $W_{ij}$ only with $d$ contiguous physical spins) obey area-law in any number of dimensions~\citep{Entanglement}, and can be represented by an MPS with a finite bond dimension $D$~[\onlinecite{MPStoRBM}]. However, some MPS states, AKLT state being an example, cannot be approximated by a finite-range RBM~\citep{MPStoRBM}. The infinite-range RBM can arbitrary well approximate any MPS, but the number of required hidden units is exponential in the bond dimension of MPS~\citep{neuralnet}, which is impractical for numerical applications. Thus, it is important to understand how well {\it shallow} RBM states, with relatively low density of hidden spins $\alpha$, which can be efficiently optimized, can approximate generic MPS states. 

Here, we provide several results regarding the performance, representation power, and the uniqueness of the  RBM ansatz. First, by connecting RBM to perturbation theory, we explain the surprisingly good performance of RBM with few variational parameters for certain models, reported in Ref.~\citep{carleo}. 
To that end, we show that RBM with $\alpha = 1$ already captures several orders of the perturbation series for those models. The connection to the perturbation theory naturally suggests an improvement for the optimisation algorithm, which we explore. Second, this observation points to a whole class of variational wave functions, similar to RBM; we demonstrate that their performance in simple models is comparable to that of RBM. Third, we investigate the performance of RBM for {\it general} local Hamiltonians, picked at random, as well as its ability to approximate random MPSs in the practically interesting case of $\alpha = 1$. By applying the ansatz to various random realisations of these systems, we show that the performance of RBM is universally determined by the entanglement entropy and range of correlations.


{\it  RBM and perturbation theory.}---  To illustrate the connection of the structure of RBM to perturbation series we first focus on a solvable, 1D transverse-field Ising (TFI) model:
\begin{equation}
\label{TFI}
H_{TFI}=-  \sum_i \sigma^x_i -J\sum_i \sigma^z_{i} \sigma^z_{i+1},
\end{equation} 
where $\sigma_i^\beta, \beta=x,y,z$ are the standard Pauli operators. 

We consider two points in the phase diagram: $J=0.5$ (paramagnetic phase), and $J=2$ (ferromagnetic phase). The GS wave function at these points can be presented in the following form, using perturbation theory:
\begin{equation}
\label{pert_exp}
|\psi\rangle = \frac{e^{\hat{W}}|\psi_0\rangle}{\langle \psi_0|e^{\hat{W}^\dagger}e^{\hat{W}}|\psi_0\rangle^{\frac{1}{2}}},
\end{equation}
where $|\psi_0\rangle$ is the unperturbed wave function: polarized along $x$-axis or $z$-axis for the paramagnetic and ferromagnetic case, respectively. The term  $\hat{W} = \sum_n\hat{W}_n$ has a perturbative structure, with the $n$-th term accounting for the $n$-th order correction in a perturbation series. The terms $\hat{W}_n$ become less and less local as $n$ is increased. Its explicit form is obtained by iteratively solving the Schr\"odinger equation 
\begin{equation}
\label{shrod}
( H  e^{\hat{W}}-e^{\hat{W}} E)|\psi_0\rangle = 0.
\end{equation}

\begin{figure}[t]
\includegraphics[scale=0.4]{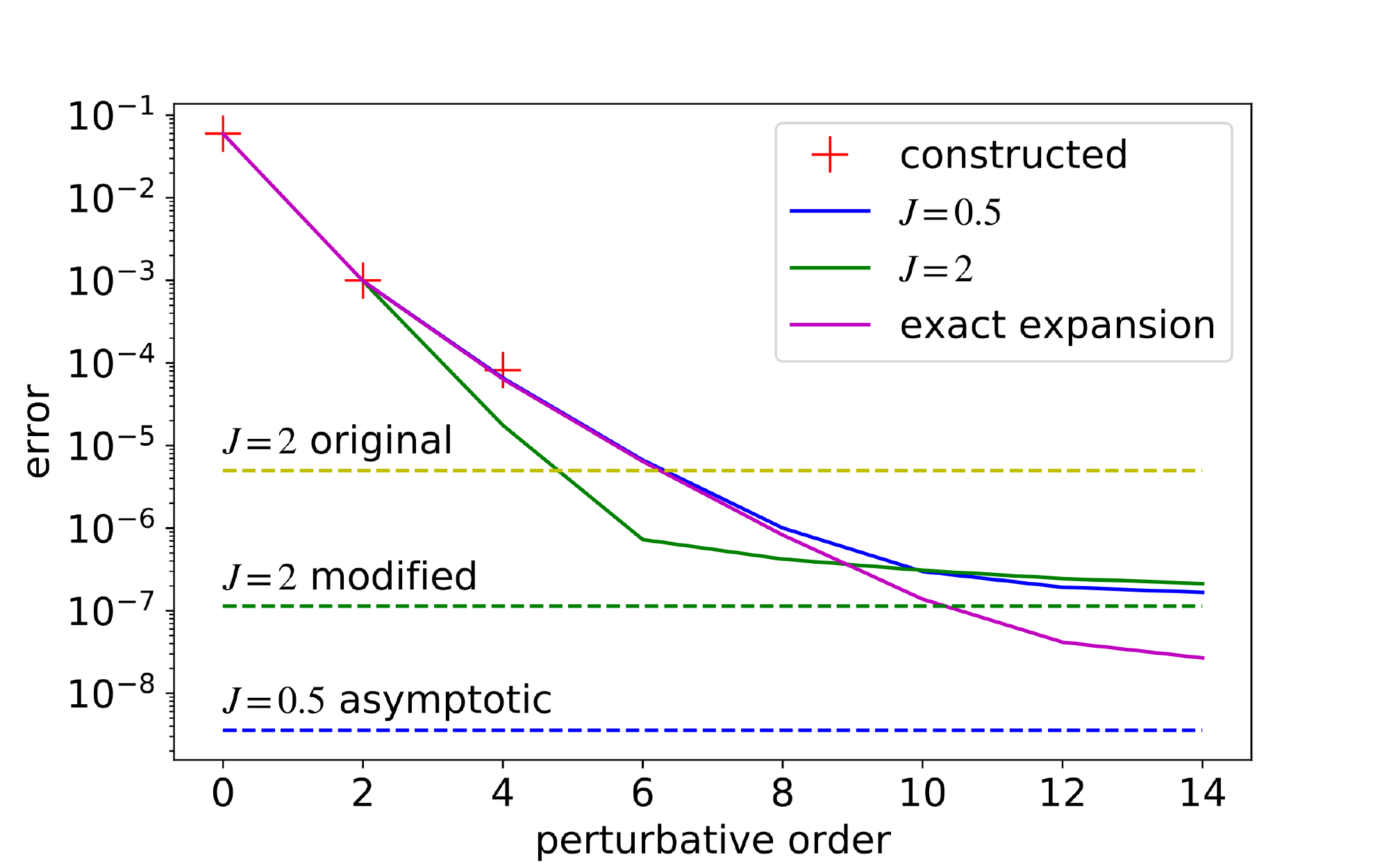}
\caption{\label{fig:er_or}
Comparison of the energy accuracy obtained by the variational wave function with finite-range, translationally invariant couplings $W_{ij}$ (blue for the paramagnetic case, green for the ferromagnetic) to the perturbative expansion of the energy computed exactly (purple), as a function of the perturbative order. 
The results for the analytically constructed RBM are denoted by crosses. The asymptotic values of error for the paramagnetic case (blue dashed line), the asymptotic value for the ferromagnetic point with a gradually increased RBM range $d$ (green dashed line), and the results of the original algorithm (yellow dashed line) are shown.	
The precision of the analytically constructed RBM wave function falls onto the curve for the exact expansion, indicating that the RBM ansatz captures several orders of perturbation theory. At higher orders the performance gain for RBM is slowing down and eventually saturates. The modified algorithm based on the perturbation theory argument outperforms the original one by 2 orders of magnitude for $J = 2$. The results are obtained for a system of $N=18$ spins.}	
\end{figure}

The RBM wave function \eqref{RBM_ans} can be rewritten in a form similar to Eq.\ \eqref{pert_exp}:
\begin{align}
\Psi_{RBM}(\{\sigma_j\}) = e^{\tilde{W}}|\tilde{\psi}_0\rangle,
\end{align}
where $\tilde{W} = \sum_j a_j \sigma^z_j+\sum^M_{i=1}\ln\cosh\left(b_i+\sum_{j}W_{ij}\sigma^z_j\right)$ and $|\tilde{\psi}_0\rangle$ stands for the state polarised along the $x$-axis.

\begin{figure*}[t]
\includegraphics[scale=0.33,width=1.99\columnwidth]{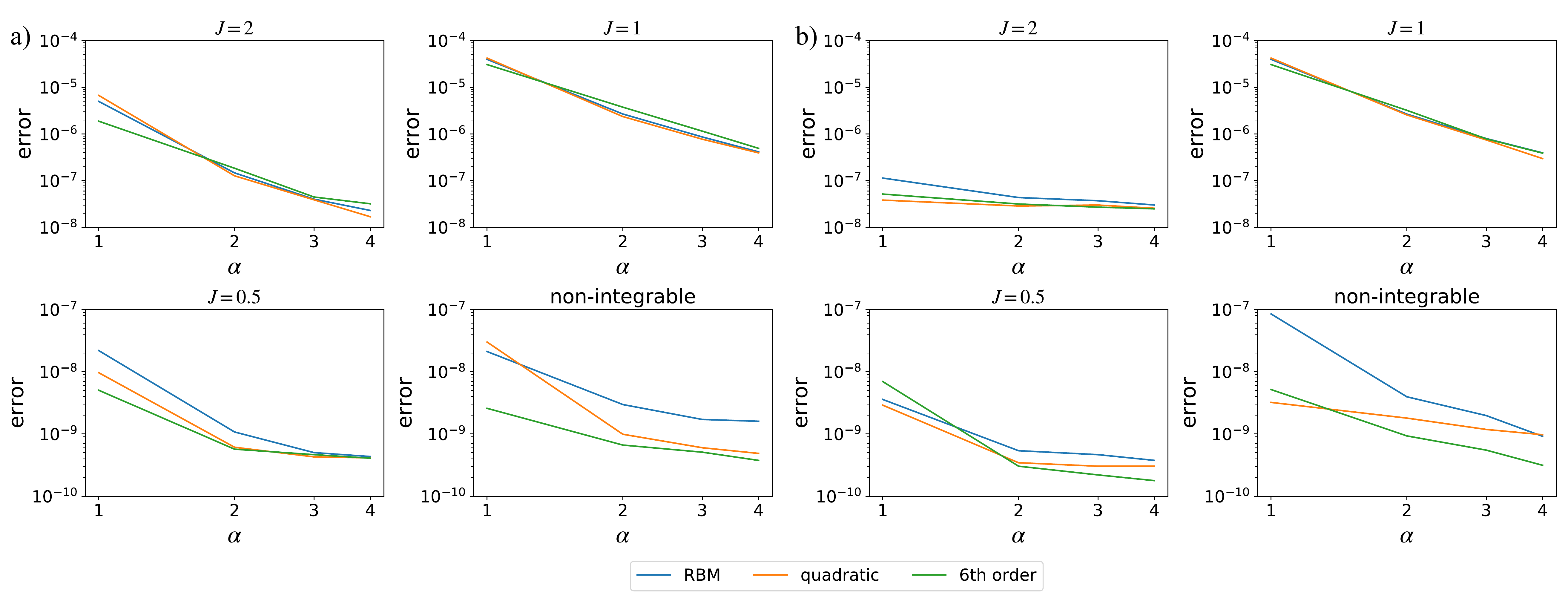}
\caption{Comparison of the performance of the RBM ansatz and two polynomial ansatze using the original algorithm (a) and the one when the range of the couplings $W_{ij}$ is gradually increased (b). Four Hamiltonians have been considered: ferromagnetic, paramagnetic, critical, and non-integrable (see main text). In all cases, the modified algorithm allows one to achieve comparable, or lower error. The system size is $N=18$.}
\label{fig:er_layer}
\end{figure*}

We note that for the paramagnetic case $\hat{W}$ can be expressed solely in terms of $\{\sigma^z_i\}$ operators 
and the $0$th order wave function $|\psi_0\rangle$ coincides with $|\tilde{\psi}_0\rangle$ (see Appendix for more details). This allows us to build a map between the perturbation series and the RBM ansatz at the paramagnetic point. By choosing $\alpha = 1$, $a_j = 0$, $b_i = 0$ and  $W_{ij} = \sum_{j'=0}^{3}\delta_{i,j+j'} \text{arctanh}\ w_{j'}$ with $w_0 = 1-J^4$, $w_1 = J-2J^3$, $w_2 = J^2+J^3$, $w_3 = 2J^3$ we are able to reproduce the first three orders of the perturbation series exactly.  

Clearly, the RBM ansatz is not capable of capturing all orders of perturbation theory. The $n$-th order term in the perturbative expansion, $\hat{W}_n$, generally contains operators with the support that scales as $O(n)$. Thus, the number of all possible terms at $n$th order scales exponentially with $n$, and exponentially many variational parameters would be needed to exactly reproduce that. However, even a finite number of terms is sufficient to approximate GS energy with a high precision. This is illustrated in Fig.\ \ref{fig:er_or}. At low orders the curve describing RBM precision of the paramagnetic GS energy approximation (blue) follows closely the expansion in powers of $J$  of the exact energy (purple), computed using Jordan-Wigner transformation. At high orders ($\geq 8$), the RBM performance becomes worse, which suggests that higher order corrections are captured only approximately.    

In Fig.\ \ref{fig:er_or} we relate a given RBM ansatz to a certain order of perturbation theory. We achieve this by taking the range $d$ of interactions $W_{ij}$ to be equal to the largest support among the irreducible terms that contribute to the corresponding order of the perturbative expansion of the energy. For example, the second order correction to the energy is determined by the first order correction to the GS wave function in the case of TFI model, which gives $d = 2$ ($d=1$) for the paramagnetic (ferromagnetic) point.

The explicit construction used above does not directly carry over to the ferromagnetic case, since in the unperturbed GS $|\psi_0\rangle$ spins are aligned along $z$-axis rather than $x$-axis and consequently $\hat{W}$ contains the spin-flip terms ($\sigma^x_i$ or $\sigma^y_i$), absent in the RBM wave function \eqref{RBM_ans}. A straightforward way to circumvent this is to rotate the basis to make the spins in the unperturbed GS polarized along the $x$-axis. In particular, this allows us to get a relative energy error $\sim\!\! 10^{-6}$ for the ferromagnetic point of 1D TFI, which is one order of magnitude smaller than the result obtained without rotation of the basis~\cite{carleo} (yellow dashed line in Fig.~\ref{fig:er_or}).

Practically, the parametric argument suggests that in RBM states with relatively low $\alpha\gtrsim 1$, there should be enough variational parameters to capture several lowest orders of the perturbative expansion. There is however a tradeoff: RBM states with larger values of $\alpha$, while forming a larger variational manifold, are harder to optimize efficiently, as one can get stuck in local minima. To enforce perturbative structure we slightly modify the original optimisation algorithm~\citep{carleo} in the following way: we gradually increase the range of variational couplings $W_{ij}$, at each step starting with the optimal parameters obtained in the previous step. 
We found that this modified procedure has a faster convergence, yielding lower error that the original procedure already at range $d=3$.

\begin{figure*}[t]
\includegraphics[scale=0.33,width=1.99\columnwidth]{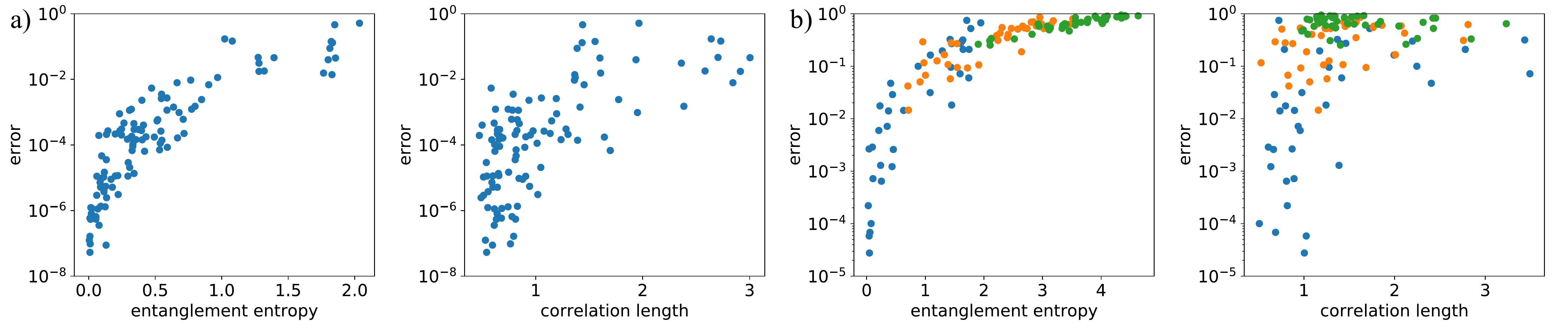}
\caption{The scatter plots of the RBM error vs the entanglement entropy of the subsystem of six cites and vs the correlation length for the problem of the GS energy approximation of random Hamiltonian (a) and for the problem of MPS approximation (b). For the latter problem we consider MPS of the bond dimensions $2$ (blue), $4$ (orange), $8$(green). The strong correlation of the RBM performance with entanglement is evident in both problems. In both cases the system size is $N = 18$.}
\label{fig:cor_ent}
\end{figure*}

{\it  A family of ans\"atze.}--- Another interesting question concerns the uniqueness of the RBM ansatz. The above discussion suggests that it belongs to a whole class of variational wave functions which can reproduce perturbative expansions. To demonstrate this, we investigate two examples below. In the first example, each hidden spin generates a quadratic polynomial:
\begin{equation}
\Psi_{quad}(\{\sigma_j\}) = \prod^M_{i=1}\left( 1 + \alpha_i \Big[b_i+\sum_{j}W_{ij}\sigma^z_j\Big]^2 \right),
\end{equation}
and coefficients $\alpha_i$ are variational parameters. The second trial ansatz is sixth-order polynomial with only even powers and fixed coefficients
\begin{equation}
\Psi_{fix}(\{\sigma_j\}) = \prod^M_{i=1}\sum_{k=0}^3\left(b_i+\sum_{j}W_{ij}\sigma^z_j\right)^{2k}.
\end{equation}
We restrict ourselves to even powers only, since we found such ans\"atze to yield a better performance.

The approximative power of these ans\"atze is summarized in the Fig.~\ref{fig:er_layer}, where we plot the accuracy of the GS energy approximation as a function of the hidden-unit density, either using the original algorithm or the modified one. The numerical results are presented for the system of $N=18$ spins and four Hamiltonians, namely, paramagnetic ($J=0.5$), ferromagnetic ($J=2$), critical ($J=1$) and a non-integrable one, defined as
\begin{equation}
H_{ni}= - \sum_i \sigma^x_i- h\sum_i \sigma^z_i  - J\sum_i \sigma^z_{i} \sigma^z_{i+1},
\end{equation}
where we choose $J=1$ and $h=0.5$. All three ans\"atze exhibit similar numerical performance, which indicates that the RBM ansatz belongs to a wider class of functions that are capable of capturing local correlations.

{\it  General Hamiltonians.}--- While TFI model provides a good test for the variational wave functions, it is important to test RBM for more general Hamiltonians. As an example, we consider a family of translationally invariant Hamiltonians with random nearest neighbour spin-spin interactions:
\begin{equation}
\label{ranham}
H_{rand}=\sum_i\sum_{\alpha,\beta = \{0,x,y,z\}} J_{\alpha\beta}\sigma^\alpha_i \sigma^\beta_{i+1},
\end{equation} 
with $J_{\alpha\beta}$ being independent random coefficients uniformly distributed in the range $[-0.5,0.5]$. We post-select Hamiltonians with translationally  invariant GS wave functions and use the RBM ansatz with $\alpha = 1$ and translationally invariant variational parameters to approximate the GS energy of these Hamiltonians.

By testing the RBM performance for different realizations of the Hamiltonian (\ref{ranham}), we discovered that the error in approximating the GS energy has a very broad distribution. To analyze the nature of this spread, we also studied the properties of the exact GS of the corresponding Hamiltonians (computed using exact diagonalization of system with $N=18$). Namely, we study how the RBM performance depends on the entanglement entropy and the correlation length~\footnote{{If $K_i(n)=\langle\sigma_1^i\sigma_{1+n}^i\rangle_c$}, then the correlation length $l$ is defined by {$l=\textrm{max}_{i=\{x,y,z\}}\sum_n \textrm{dist} (1,1+n) |K_i(n)|/\sum_n |K_i(n)|$}, where {$\textrm{dist}(i,j)$} is the distance between to spins at sites $i,j$ with the periodically boundary conditions being taken into account} (Fig.\ \ref{fig:cor_ent}). The positive correlation of the error with both quantities is evident, indicating that RBM performs best for states with short-ranged correlations and low entanglement, as expected.

{\it RBM and MPS.}--- We can further understand the interplay of RBM and entanglement by studying the relation of the ansatz to MPSs. Let us first provide an exact and transparent mapping from finite-range RBM to MPS in 1D. The RBM ansatz with coupling range $d$ can be written as
\begin{equation}
\label{productform}
\Psi_{RBM}(\{\sigma_j\}) = \prod_i f^{(i)}_{\sigma_i..\sigma_{i+d-1}},
\end{equation}
where $f^{(i)}_{\sigma_i..\sigma_{i+d-1}}$ is a function acting on $d$ neighbouring spins. For example, to write the RBM wave function with $\alpha=1$ in this form, we can choose $f^{(i)}_{\sigma_i..\sigma_{i+d-1}} = e^{a_i \sigma_i^z}\cosh(b_i+\sum^{d-1}_{j=0}W_{i,i+j}\sigma^z_{i+j})$. Next, we construct a mapping from this form to an MPS form. We can associate the tensor $A^{\sigma_i}_{i-d,..,i+d-1}$ with each function $f^{(i)}_{\sigma_i..\sigma_{i+d-1}}$ using the following rule
\begin{equation}
\label{mapp}
A^{\sigma_i}_{j_{1}..j_{d-1}k_{1}..k_{d-1}} = f^{(i)}_{\sigma_i k_{1}..k_{d-1}}\delta_{\sigma_ij_1}\delta_{k_1j_2}\cdot..\cdot\delta_{k_{d-2}j_{d-1}},
\end{equation}
where the first $d-1$ indices ($j_{1}..j_{d-1}$) of the tensor should be considered as one index taking $2^{d-1}$ values, which should be contracted with the $(i-1)$th tensor and the last $d-1$ indices ($k_{1}..k_{d-1}$) are contracted with the $(i+1)$th tensor. Then the RBM-wave function can be written in the form
\begin{equation}
\Psi_{RBM}=\sum_{\{\sigma_j\}} Tr(A^{\sigma_1}..A^{\sigma_N})|\{\sigma_j\}\rangle
\end{equation}
Our argument makes use of the fact that the spin system either has periodic boundary conditions or is infinite, however, this argument can be effortlessly generalized to finite size systems with boundaries. Also the slightly modified argument relates RBM to MPS in higher dimensions. 

In contrast to Ref.~[\onlinecite{MPStoRBM}], where the mapping was made to an MPS with a bond dimension $D$ that is exponential in the number of pairs of the connected hidden unit and the physical spin that have a given spin in between ($\sim~\!\!e^{d^2}$), we get an exponential ($D \sim e^d$) scaling with the range of RBM. 

To study the inverse mapping, we use RBM with $\alpha =1$ to approximate MPSs with a given bond dimension. We generate random translationally invariant MPSs $|\Psi_{MPS}\rangle$ with $D=2,4,8$ for a system of $N=18$ spins~\footnote{We construct MPSs as products of identical random matrices that have independent random entries with both real and imaginary parts being uniformly distributed in the interval $[-0.5,0.5]$.}. Then we minimize the quantity $1-|\langle \Psi_{MPS}|\Psi_{RBM}\rangle|^2$, which is plotted against the entanglement entropy and the correlation length in Fig.\ \ref{fig:cor_ent}. One can see a clear dependence of the error on the entanglement entropy, which is in agreement with the results obtained for random Hamiltonians above.

{\it  Conclusions.}--- In this paper we considered the RBM ``black box" from the physical prospective. We pointed out an explicit connection between the RBM ansatz and perturbative description of ground states of gapped models, which explains the remarkable accuracy of few-layer RBM for simple models such as the TFI model.~[\onlinecite{carleo}]. In some cases, several orders of perturbation series can be exactly captured by the RBM wave function. Even when this is not the case, the mere existence of the perturbative series allowed us to introduce a simple modification of the optimization algorithm of RBM with an improved performance. Furthermore, the intuitive connection with the perturbation theory helped us to introduce a whole family of ans\"atze that have a performance similar to that of RBM. This suggests that RBM is not a unique ansatz, but rather belongs to a broad class of similarly powerful variational wave functions. 

Finally, we have investigated the performance of practical, few-layer RBM states, in approximating ground states of {\it random} (but translationally invariant) local spin Hamiltonians. We found that the approximation error has a broad distribution, and the success of few-layer RBM is mostly determined by the degree of entanglement in the ground state. 
It is natural to expect that states with low entanglement can be well approximated using just few orders of perturbation theory, which RBM can capture. 

We leave for the future work the detailed investigation of how well infinite-range RBM can reproduce critical many-body states. Another open question is whether the volume-law entanglement that RBM states generally have can be useful for approximating, e.g., excited states and entanglement spreading in many-body systems. 

{\it  Acknowledgement.}--- This work was supported by the Swiss National Science Foundation. We thank G.~Carleo for helpful correspondence. The numerical simulations were performed on the HPC cluster Baobab.

\vspace{1cm}

\onecolumngrid
\appendix
\section{Appendix: An exact mapping of the third-order perturbative wave function to the RBM ansatz}
\label{appx:3ord}

As it is stated in the main text (Eq. (\ref{pert_exp})) $\hat{W}$ is determined by

\begin{equation}
( H  e^{\hat{W}}-e^{\hat{W}} E)|\psi_0\rangle = 0.
\end{equation}
Let us write $\hat{W}$ and the energy $E$ in the form $\hat{W} = \hat{W}_1+\hat{W}_2+\hat{W}_3+O(J^4)$ and $E=E_0+E_1+E_2+E_3+O(J^4)$ to outline their perturbative structure. 

To simplify analysis we want to fix the freedom in choice of $\hat{W}$. It could be chosen to depend solely on $\{\sigma^z_i\}$. This is the case, since an action of any operator on $|\psi_0\rangle$ that is the product of $\{\sigma^\alpha_i\}$ results only in spin flips at certain positions. Thus, to cancel such terms we need to put $\sigma^z_i$ in the corresponding positions of $\hat{W}$ . When commuted with $H_0$, some $\sigma^z_i$ would be changed to  $\sigma^y_i$, but the resulting action on $|\psi_0\rangle$ stays the same up to a phase ($\sigma^z_i|\psi
_0\rangle=i\sigma^y_i|\psi_0\rangle$). 

Two important simplifications follow from this observation . Namely that $[V,\hat{W}]=0$ at any order and $e^{\hat{W}_n+\hat{W}_m}=e^{\hat{W}_n}e^{\hat{W}_m}$. Also, a simple property of the TFI Hamiltonian is that all odd energy corrections are zero. Having said these the equations that determine the terms of interest in $\hat{W}$ can be written as
\begin{align}
&\left([H_0,\hat{W_1}] + V \right)|\psi_0\rangle = 0,\\
&\left([H_0,\hat{W_2}] + \frac{1}{2}[[H_0,\hat{W_1}],\hat{W_1}]\right)|\psi_0\rangle = E_2|\psi_0\rangle,\\
&\left([H_0,\hat{W_3}] + [[H_0,\hat{W_1}],\hat{W_2}]+ \frac{1}{6}[[[H_0,\hat{W_1}],\hat{W_1}],\hat{W_1}] \right)|\psi_0\rangle = 0.
\end{align}
To have more compact notation we denote $\lambda\equiv \frac{J}{4}$. If we take $\hat{W}_1 = \lambda\sum_i \sigma^z_i\sigma^z_{i+1}$, then $[H_0,\hat{W_1}] = 2i\lambda(\sum_i \sigma^y_i\sigma^z_{i+1}+\sum_i \sigma^z_i\sigma^y_{i+1})$. Therefore,the term that should be cancelled in the second order is 
\begin{equation}
\frac{1}{2}[[H_0,\hat{W_1}],\hat{W_1}] = -4\lambda^2\sum_i (\sigma^x_i + \sigma^z_i \sigma^x_{i+1} \sigma^z_{i+2} ).
\end{equation}
We choose $\hat{W}_2 = \lambda^2\sum_i \sigma^z_i\sigma^z_{i+2}$ and get an extensive energy correction $E_2 = -4\lambda^2\sum_i 1$, then $[H_0,\hat{W_2}] = 2i\lambda^2(\sum_i \sigma^y_i\sigma^z_{i+2}+\sum_i \sigma^z_i\sigma^y_{i+2})$ . The third order interaction is then given by
\begin{eqnarray}
[[H_0,\hat{W_1}],\hat{W_2}]+ \frac{1}{6}[[[H_0,\hat{W_1}],\hat{W_1}],\hat{W_1}] =\frac{16\lambda^3i}{3}\sum_i\left(\sigma^y_i\sigma^z_{i+1}+\sigma^z_i\sigma^y_{i+1}\right) - \\
4\lambda^3\sum_i\left(\sigma^x_i\sigma^z_{i+1}\sigma^z_{i+2}+\sigma^z_i\sigma^z_{i+1}\sigma^y_{i+2}+\sigma^z_i\sigma^x_{i+1}\sigma^z_{i+3}+\sigma^z_i\sigma^x_{i+2}\sigma^z_{i+3}\right).
\end{eqnarray}
This gives $\hat{W}_3= 2\lambda^3\sum_i(\sigma^z_i\sigma^z_{i+3}+ \sigma^z_i\sigma^z_{i+1}/3)$ and the total wave function up to normalization could be written as
\begin{equation}
\label{pert_func}
|\psi\rangle = e^{\hat{W}}|\psi_0\rangle + O(\lambda^4),
\end{equation}
with
\begin{equation}
\hat{W} = \sum_i\left((\lambda-\frac{2\lambda^3}{3})\sigma^z_i\sigma^z_{i+1}+\lambda^2\sigma^z_i\sigma^z_{i+2}+2\lambda^3\sigma^z_i\sigma^z_{i+3}\right).
\end{equation}

At this point is is clear how RBM ansatz could be constructed in order to recover perturbative expansion, i.e. Taylor expansion of each cosine should reproduce local terms in the exponent of  \ref{pert_func}. Since we need terms only even in operators $\sigma^z$, it is natural to take parameters $a_i$, $b_j$ of RBM ansatz to zero. The parameters $W_{ij}$ are chosen to be translationally invariant, i.e. dependent only on difference between the position of a hidden-unit and the physical spin $W_{ij}=\tilde{W}_{i-j}$. Since the largest support of the terms in $\hat{W}$ is four, we restrict ourselves to four non-zero parameters $\tilde{W}_{i}$, $\{i=0,..,3\}$. Applying all the assumptions to the RBM wave function (Eq.(\ref{RBM_ans})), we may write ansatz in the form
\begin{align}
\Psi_{RBM}(\{\sigma_j\})=\left(\prod_{i=0}^3\cosh \tilde{W}_i\right)^N\prod_j \left(1 + w_0 w_1 \sigma_j^z\sigma_{j+1}^z+ w_0 w_2 \sigma_j^z\sigma_{j+2}^z+ w_0 w_3 \sigma_j^z\sigma_{j+3}^z + w_1 w_2 \sigma_{j+1}^z\sigma_{j+2}^z+\right.\\ \nonumber
\left. + w_1 w_3 \sigma_{j+1}^z\sigma_{j+3}^z+ w_2 w_3 \sigma_{j+2}^z\sigma_{j+3}^z+ w_0 w_1 w_2 w_3\sigma_j^z\sigma_{j+1}^z\sigma_{j+2}^z\sigma_{j+3}^z\right),
\end{align}
where $w_i = \tanh \tilde{W_i}$. Since we assume this expression to have the perturbative structure , the terms containing Pauli matrices should be small. Thus, we can  take approximate logarithm of this expression to get the exponential form as of Eq. \ref{pert_func} up to normalization. Then we could easily identify the value of parameters $w_i$ that reproduce the perturbative expansion to the third order: $w_0 = 1$ (it is possible with an exponential precision for finite $\tilde{W}_0$ ), $w_1 = \lambda -2\lambda^3$, $w_2 = \lambda^2 +\lambda^3$ and $w_3 = 2\lambda^3$.


\bibliographystyle{apsrev4-1}
\bibliography{citation}

\end{document}